# Estimating Reaction Rate Parameters in Cell Signaling Pathways Using Extreme Decomposition and Belief Propagation Tailored for Data-Rich Cases

Tri Hieu Nim[1], Le Luo[2,3], Marie-Véronique Clément[2,3], Jacob K. White[1,4], Lisa Tucker-Kellogg[1,5] *,

[1]Computational Systems Biology Programme, Singapore-MIT Alliance, Singapore 117576

[2]NUS Graduate School for Integrative Sciences, National University of Singapore, Singapore 119260

[3]Department of Biochemistry, Yong Loo Lin School of Medicine, National University of Singapore, Singapore 119260

[4]Department of Electrical Engineering and Computer Science, MIT, Cambridge, MA, USA 02139

[5]Mechanobiology Institute, National University of Singapore, Singapore 117411

*Correspondence to LisaTK@nus.edu.sg

**ABSTRACT**

**Motivation:** Modeling biological signaling networks using ordinary differential equations (ODEs) has proven to be a powerful technique for generating insight into cellular dynamics, but it typically requires estimating rate parameters based on experimentally observed concentrations. New measurement methods can measure concentrations for all molecular species in a pathway, which creates a new opportunity to decompose the optimization of rate parameters.

**Results:** In contrast with conventional methods that minimize the disagreement between simulated and observed concentrations, the BPPE method fits a spline curve through the observed concentration points, and then matches the derivatives of the spline-curve to the production and consumption of each species. Whereas traditional methods follow the ODEs exactly and then attempt to match the data, BPPE follows the data exactly and then attempts to match the ODEs. The new objective function is an extreme decomposition of the problem because each factor in the function is enforcing the equality of one ODE at one timeslice. A "loopy belief propagation" algorithm solves this factorized approximation of the parameter estimation problem providing systematic coverage of the search space and unique asymptotic behavior; the run time is polynomial in the number of molecules and timepoints, but exponential in the degree of the biochemical network. The implementation is a global-local hybrid optimization, and we compare with the performance of local, global, and hybrid methods. BPPE is demonstrated for a novel model of Akt activation dynamics including redox-mediated inactivation of PTEN.

**Availability:** Software and supplementary information are available at http://webbppe.nus.edu.sg:8080/opal2/WebBPPE

**Contact:** LisaTK@nus.edu.sg

**Keywords**: probabilistic graphical models, physico-chemical modeling, systems biology, signal transduction.

## 1 INTRODUCTION

Ordinary differential equation (ODE) models have become popular for describing the biochemical pathways that govern signal transduction dynamics in cells [1-4]. In this ODE framework, the chemical kinetic rate constants, or the coefficients of the differential equations, are typically not determined by direct experiments. These rates are then unknown parameters that must be estimated, generally by regression or fitting the global behavior of the simulated model to the experimentally observed concentrations. Estimating the rate parameters is a difficult problem due to slow convergence, stagnation, or convergence to non-optimal local minima [5]. The kinetic rate constants are crucial for the dynamic behavior of the effects being simulated, and many methods of parameter estimation have been proposed.

The rate parameter estimation problem can naturally be formulated as minimizing a sum of squared errors, where each error is a difference between simulated concentration and observed concentration, and the summation is over time points and/or experimental treatments. A wide variety of optimization methods [6-11] are applicable to such problems, and can be grouped into local, global, and hybrid search methods. Local searches typically use nonlinear least squares optimization, and run very quickly, but they only explore the parameter space around a given start value. Therefore, local methods depend on a good initial guess, meaning an initial guess that is within the basin of convergence of the global optimum. If the objective function is rugged with many local optima, the chance of initializing in the optimal basin exponentially in the number of dimensions. Global methods [12] typically use heuristics and random sampling to explore the entire domain of parameters, and the run time depends on the level of sampling performed. Despite the "global" terminology, the results produced by such methods carry no guarantee of global optimality, except theoretically when the sampling time goes to infinity. Many local and global optimization methods have a subclass of problems (number of parameters, amount of data, amount of noise, manner of initialization, etc.) for which a particular method performs best. In other scientific areas where optimization is required, hybrid global-local methods have performed well, and hybrid methods have recently become popular due to performance features that combine high speed with broad coverage [13-17].

Decomposition is another common strategy for large-scale nonlinear optimization, including rate constant estimation. Various methods of decomposing the parameter estimation problem pro-vide important precedent for the current work. Koh et al. established formal conditions for decomposing a network based on cutting the network topology at molecule nodes that have observed concentrations [18]. The resulting sub-pathways could be estimated independently with traditional nonlinear optimization, although some difficulties remained for re-assembling the local results into a global solution. Their subsequent works performs parameter estimation using probabilities to encode a distribution of parameter estimates, and this approach permits new experimental data points to be utilized by an online algorithm with probabilistic inference in a factor graph [19, 20]. Quach et al. performed a form of decomposition that permitted Kalman filters to be used for parameter estimation [21]. Chou et al. devised a method that completely decomposes the network into single ODEs, akin to our factorization. Their implementation expected the derivatives of all molecular concentrations to be provided [22], and the resulting optimization problem was approached using an alternating regression method which guessed some initial parameter values and then sought revised estimates in a manner similar to the EM algorithm. Another recent method also decomposes the network into single ODEs [23] but the parameter space is still explored by random sampling and generating simulated trajectories. The concept of decomposition is an important concept for attacking very high-dimensional problems.

Belief propagation (BP) is an algorithm for probabilistic inference in graphical models such as factor graphs, Bayesian belief networks, or Markov random fields [24]. Brute force methods to infer probabilities algebraically would have exponential run time in the number of variables, but BP runs in polynomial time by exploiting the locality and sparsity of the graph to achieve global inference using only local "message passing" operations [24]. Another way to summarize the innovation of BP is that it exploits the conditional independence in a network to define local sub-problems, and to propagate the solutions of the sub-problems to only the sub-problems that require the information. Belief Propagation has been applied to many problems in biological data analysis and biological system modeling. A major body of work involves the application of belief propagation to gene expression analysis [25]. For pathway modeling, belief propagation has been used for learning regulatory interactions between genes [26], and more recent work has used it for parameter estimation or reconciling alternative solutions [19, 20].

In this paper, we describe a novel transformation of the reaction rate parameter estimation problem into an approximate "dual" problem with factorized objective function. The novel objective function can be optimized heuristically by a deterministic global inference algorithm, loopy belief propagation. While sharing the same framework for decomposition with the work by Koh et al. [19, 20] and Chou et al. [22], this work diverges by employing search space discretization for pre-computation of ODE violations with derivative approximation, and voxel sampling that allows efficient execution of loopy belief propagation. Our approach is accompanied by different performance trade-offs, different data requirements, and a different asymptotic run time. Remarkably, the run time of the method is polynomial in the number of variable parameters, number of molecules, and number of timepoints, but it is exponential in the degree of the signaling pathway. Conventional methods search a space that is exponential in the number of parameters. The price for the improved scalability of our method is the requirement that concentration data be provided for all molecules in the pathway.

Finally, we show the practical performance of this. We compare its performance with a variety of local, global, and hybrid methods for artificially-generated cases with uniformly increasing size, and for a realistic biological problem. We build a model of Akt activation dynamics, including our recent experimental finding that the PTEN phosphatase can be inactivated by a redox-mediated post-translational modification, such as S-nitrosylation [27]. In this pathway, our parameter estimation method showed superior performance over all other methods being investigated, although for the same system all methods showed dramatic (and equivalent) degradation of performance when artificial noise was introduced into the system. This example shows that that landscape of estimation tasks is rugged with small changes in problem inputs carrying large changes in performances, meaning that keeping a variety of methods can be useful. In practice, the empirical performance of this method is competitive with other methods in large-scale testing, indicating that this approach may provide useful ingredients for future work.

## 2 PRELIMINARIES

### 2.1 Ordinary differential equation (ODE) for mass action kinetics (MAK)

Mass action kinetics (MAK) explains the behavior of solutions by adding the rates of the elementary reactions, each weighted by a corresponding set of stoichiometric coefficients, sometimes called the reaction rates or the kinetic constants. Elementary reactions provide a framework for constructing ordinary differential equations (ODEs) to be satisfied by the time-evolved concentrations in the solution. For example consider a 4-species artificial pathway whose reactions can be described in chemical notation as $A + B \rightarrow C \rightarrow D$. Using $k_1$ and $k_2$ to denote the left and right reaction rates, respectively and $\vec{x}$ to denote the set of all species in the MAK system, we can represent the same system in ODEs as below:

$$\frac{d}{dt}A(t,\vec{k}) = f_1\left(\vec{x}(t,\vec{k}),\vec{k}\right) = -k_1 A(t,\vec{k})B(t,\vec{k})$$

$$\frac{d}{dt}B(t,\vec{k}) = f_2\left(\vec{x}(t,\vec{k}),\vec{k}\right) = -k_1 A(t,\vec{k})B(t,\vec{k})$$

$$\frac{d}{dt}C(t,\vec{k}) = f_3\left(\vec{x}(t,\vec{k}),\vec{k}\right) = k_1 A(t,\vec{k})B(t,\vec{k}) - k_2 C(t,\vec{k})$$

$$\frac{d}{dt}D(t,\vec{k}) = f_4\left(\vec{x}(t,\vec{k}),\vec{k}\right) = k_2 C(t,\vec{k})$$

The above equations are a model of the system and they specify the time evolution of each species in the system, dependent on time $t$ and rate parameter vector $\vec{k}$. The functions specifying these time evolutions depend on the species concentrations and the rate parameters. In the context of ODEs for MAK, rate constant estimation is essentially a nonlinear optimization problem [5]. The degree of each ODE is defined as the number of reactions, or terms, on its right hand side, analogous to the node degree in a biochemical network diagram. The degree of the system is defined as the maximum degree of all ODEs.

## 2.2 Rate constant estimation objective

To estimate the rate constants, the most standard approach is a nonlinear least square technique to minimize the weighted sum of squared error (SSE) objective function:

$$\min_{\vec{k}} \sum_{i \in \text{Species}} \frac{N_t}{\sum_{j \in \text{Timepoints}} \left(x_i^{data}(t_j)\right)^2} \sum_{j \in \text{Timepoints}} \left(x_i^{data}(t_j) - x_i^{sim}(t_j, \vec{k})\right)^2 \quad (1)$$

Here $N_t$ is the number of timepoints, $x_i^{data}(t_j)$ refers to the observed data, and $x_i^{sim}(t_j, \vec{k})$ refers to the simulated dynamics for species $x_i$ based on a set of rate constants $\vec{k}$. Local optimization methods include Levenberg-Marquardt (LM) [28] and Steepest Descent (SD) [6]. Common global methods include "Evolution Strategy using Stochastic Ranking" (SRES) [9], Genetic Algorithms (GA) [11] and Particle Swarm Optimization (PSO) [7]. As the parameter space is exponential with respect to the number of parameters, the optimization performance can degrade quickly with an increasing number of parameters. These methods are often unacceptable for parameter estimation in large biological pathways, and indeed many high-impact models continue to be built without automating the parameter estimation process [29-31]. Meanwhile innovative reaction rate estimation algorithms continue to be developed with significant improvement over previous methods, but they remain tied to a randomized sampling of an exponential-sized parameter space, and they remain under threat of intractability for larger networks.

## 3 ALTERNATIVE OBJECTIVE FUNCTION

Inspired by primal-dual transformations from the field of linear programming, we construct an alternative parameter estimation problem that resembles a "dual" of the conventional problem. The standard "primal" formulation of the problem is to optimize the agreement between model concentrations and experimental concentrations, subject to the constraint that the model concentrations (obtained by solving the ODE numerically) satisfy the ODE equations (i.e., that the concentrations are taken from a time-evolved trajectory of the ODEs). For the "dual" problem, we seek to minimize the violation of the ODEs, the amount by which the right hand side differs from the left hand side, subject to the constraint that the model concentrations exactly match the experimentally observed concentrations. In other words, we build a model based initially on the known concentrations and then attempt to match the derivatives to the reaction equations. We propose no analogous duality theorem for this primal-dual transformation, but merely the observation that a correct solution to one problem will also be a solution to the other.

## 3.1 The terms of the objective function enforce agreement with the ODEs

Given the optimization goal of maximizing the agreement between the interpolated concentration curves from observed data and the ODEs, we next seek a mathematical statement of the full objective function. Instead of driving the simulated dynamics towards observed data, the proposed method drives the interpolation of the observed data towards the required ODEs. Hence, the optimization formula is to minimize $\left| f_i\left(\vec{x}^{data}(t_j), \vec{k}\right) - \frac{\hat{d}}{dt} x_i \bigg|_{t=t_j} \right|$ for all functions $f_i$ in the ODEs. Note that this term depends only on the rate parameters and observed concentrations with no simulated concentrations. In the optimization formula, $\frac{\hat{d}}{dt} x_i \bigg|_{t=t_j}$, the approximate derivative for species $x_i$ at time $t_j$, is approximated numerically from

$$f_i\left(\vec{x}^{data}(t_{j-1}), \vec{k}\right), \vec{x}^{data}(t_{j-1}), \vec{x}^{data}(t_j), \vec{x}^{data}(t_{j+1}), f_i\left(\vec{x}^{data}(t_{j+1}), \vec{k}\right)$$

and $\vec{k}$ using clamped cubic spline interpolation (see section 3.2).

Most objective functions for parameter estimation employ the simulated concentrations from a time-evolved trajectory of the system, but simulations require all rate constants and would create a high dimensional problem. Utilizing only a small subset of the rate constants in each term and eschewing simulation is the innovation that permits extreme decomposition, as shown later. Note that the approximation of derivatives will impose some inaccuracy in the computation, to be analyzed empirically in the results section.

## 3.2 A clamped cubic spline approximates the derivatives

Splines can provide smooth interpolation among discrete data points, allowing the slope of a curve to be estimated from a series of discrete timepoints. To approximate the derivatives we use clamped cubic spline interpolation [32] which solves a system of linear equations for the coefficients of a spline polynomial. Other schemes such as backwards difference and natural splines approximate the first derivatives with a higher level of error with the same number of data points. To compute splines for all species in the system, we require experimental data to be available for the concentrations of every species in the system. Hence the use of this method is restricted to data-rich cases, such as SILAC [33] datasets, proteomic measurements, or small well-studied pathways with non-proteomic measurements. Incomplete data decreases accuracy but does not preclude the use of this method as the endpoint derivatives don't have to be adjacent to the timepoint to be computed, i.e. $t_{j-2}$ may be used instead of $t_{j-1}$ when computing $\frac{\hat{d}}{dt} x_i \bigg|_{t=t_j}$.

Clamped spline interpolation approximates the first derivatives using three data points and two endpoint derivatives with error $\mathcal{O}(h^4)$, where $h$ is the interval between the data points [34]. Endpoint derivatives make the computation of the spline unique. We estimated the endpoint derivatives by computing $f_i\left(\vec{x}^{data}(t_{j\pm1}), \vec{k}\right)$, the right hand side of the ODE at the corresponding timepoints, but other methods would also be possible.

Figure 1 illustrates the derivative for species $x_i$ at timepoint $t_j$, approximated (dashed arrow) using a clamped cubic spline based on observed data for $t_{j-1}$, $t_j$ and $t_{j+1}$; and endpoint derivatives from the ODEs at timepoints $t_{j-1}$ and $t_{j+1}$.

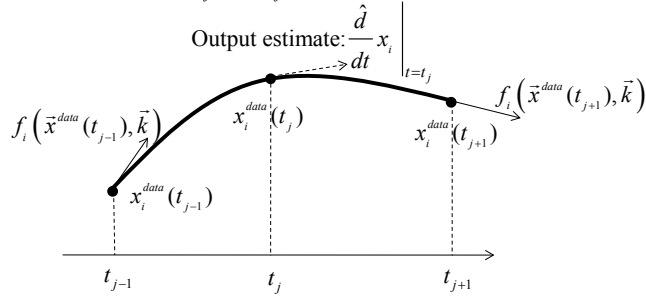

Figure 1. Illustration of computing the approximate derivative using clamped spline interpolation

We reiterate that this computation requires observed data to be available for all species so that derivatives can be estimated for all species.

### 3.3 Product of functions

The objective function, the function to be minimized for matching the ODEs, can be composed from individual terms. Among different schemes of constructing the objective function (e.g. sum of squares), we propose the product form because it is a decomposable expression, and it permits the use of probabilistic inference methods based on factor graphs. Using this product of functions (POF) objective function, the optimization problem becomes:

$$\text{POF} = \min_{\vec{k}} \prod_{i \in \text{Species}} \prod_{j \in \text{Timepoints}} \left| \left( f_i\left(\vec{x}^{data}(t_j), \vec{k}\right) - \frac{\hat{d}}{dt} x_i \Big|_{t=t_j} \right) \right| \quad (2)$$

The POF objective function lays the basis for the "extreme decomposition" approach in our method. From (2), we observe that the product of factors is minimized when each individual POF factor is minimized. As each individual ODE typically contains only a few reactions, (assuming a low-degree network) each POF term involves a small subset of the rate parameters $\vec{k}$. However, as the factors are not independent, the dependency between factors has to be resolved.

Corresponding to the factorized objective, we construct a graphical model of interdependency (section 4.1): a graph of factor nodes and parameter variable nodes, such that a factor node is connected to a variable node if and only if the variable appears in the factor. This defines a canonical factor graph, and the variables that best satisfy the factorized objective function can be inferred by running the "belief propagation" message passing algorithm on the graph [24, 35], provided the graph is acyclic. Thus we have transformed the high-dimensional optimization problem into many easy low-dimensional problems, one for each ODE at each timestep, and these many small problems yield parameter choices which can be reconciled efficiently via message passing. The only remaining obstacle is that the factor graphs in our case would generally have cycles, and standard belief propagation does not apply.

## 4 LOOPY BELIEF PROPAGATION

We now describe a method to minimize the POF objective function approximately but deterministically using Loopy Belief Propagation (LBP) on a factor graph. LBP is an approximate and heuristic version of the belief propagation algorithm [24] for computing the best-scoring values (the maximum *a posteriori* values) in a factor graph with cycles [36]. Section 4.1 describes the factor graph, defined by the POF factors and the variable parameters that are associated with each POF factor. Section 4.2 describes how a joint distribution "lookup table" addresses the local parameter optimization problem for each POF factor. Section 4.3 describes the message-passing algorithm for computing the maximum a posteriori values of the variable parameters (the optimum of the POF objective) by using the factor graph and its joint distribution tables. Section 4.4 is a brief asymptotic analysis of the algorithm, showing that the size of each joint lookup table is exponential in the degree of the graph, but the remainder of the algorithm is polynomial.

### 4.1 Factor graph

We now specify how a factor graph, which can be minimized using LBP, is defined to represent the POF objective function (Equation 1 in section 3.3). Figure 2 show a portion of the factor graph corresponding to the 4-species $A + B \rightarrow C \rightarrow D$ system at timepoint $t_j$, with variable nodes represented by circles and factor nodes by rectangles. This results in a factor graph representation of the system of ODE which represents the unknown variables and the relationship among them to be reinforced. The circles represent variable nodes, which are the unobserved parameters to be optimized. The rectangles represent the factor nodes, each of which is designed to enforce one term of the POF objective function.

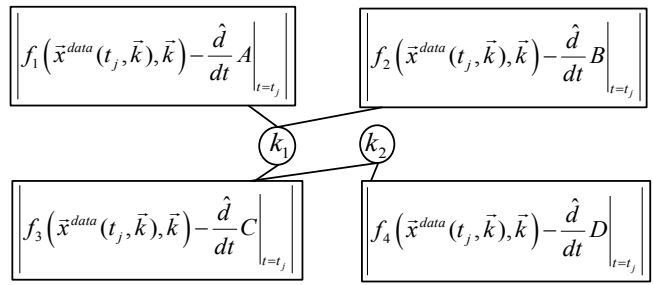

Figure 2. Factor graph representation of the 4-species system

The links (edges) of the factor graph indicate probabilistic dependence between the parameter and the factor. Sparsity of the graph shows conditional independence. Each variable node is associated with a probability distribution, continually updated during the message-passing process, to reflect which value for the variable parameter is most likely. Each factor node represents one term of the POF objective, which we recall is the level of discrepancy between left hand side of the ODE (as interpolated from data) and the right hand side of the ODE (computed using the variable parameters). The possible combinations of parameters can be represented as a joint probability distribution.

The degree of a factor node is defined as the number of variable nodes adjacent to it. In the case illustrated in Figure 2, factor node

$\left| f_3 \left( \vec{x}^{\,data}(t_j), \vec{k} \right) - \left. \dfrac{\hat{d}}{dt} C \right|_{t=t_j} \right|$ has degree 2, while all other factor nodes have degree 1. We denote $N(X)$ as the set of neighbors, or adjacent nodes, to a node $X$, which can be a variable node or a factor node.

## 4.2    Joint probability table and discretization

The true set of possible values for each rate parameter is a continuous interval of real numbers which cannot be evaluated individually. Therefore we discretize the possibilities into discrete intervals, called bins, with coarser yielding faster run time at the expense of accuracy. Each bin is represented by its midpoint (and nearby values belonging to the same bin will be represented by the same midpoint. The distribution of values for each variable is therefore a discrete distribution while the distribution of each factor node is a joint distribution. This can be considered as a "lookup joint table", which allows quick computation of the relationship among the variables. The joint probability table is computed by evaluating $\left| f_i \left( \vec{x}^{\,data}(t_j), \vec{k} \right) - \left. \dfrac{\hat{d}}{dt} x_i \right|_{t=t_j} \right|$ based on all combination of $k_i$'s, as illustrated in Figure 3.

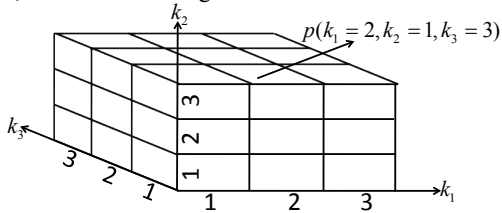

Figure 3. Illustration of joint probability table of 3 dimensions corresponding to rate constants $k_1$, $k_2$ and $k_3$.

Each dimension of a lookup table corresponds to one associated variable parameter (variable node in the factor graph). The value computed at each cell of the table (e.g., the cell with $k_1= i^{th}$ bin, $k_2=j^{th}$ bin, …) is the evaluation of the POF factor for that interval, meaning that that $f_i()$ expressions for production and consumption are instantiated using the midpoint of the $i^{th}$ bin for each occurrence of parameter $k_1$, the midpoint of the $j^{th}$ bin for parameter $k_2$, etc. When all the entries of a table have been computed, discretized minimization of a single POF factor can be performed simply by scanning the joint table to identify the minimum entry.

Due to the discretization required by this method, the output estimate for each parameter represents a range rather than a single value. Although discretization sacrifices some accuracy (analogous to round-off error), we choose the variable parameters to be discretized, sometime quite coarsely, because the output of such a method might be ideal input for a local search method, such as LM or SD, to refine using a more precise simulation-based objective function.

One way to view the innovation of this approach is that the global relationship between all the variable nodes can be approximated and pre-computed in the form of joint lookup tables -- many tractable and sparsely interconnected lookup tables. The algorithm will exploit this extreme global decomposition to choose a good neighborhood of the high-dimensional parameter space. Beyond that, the choice of a particular point in the neighborhood is a problem better suited to existing optimization methods.

## 4.3    Algorithm

Message passing algorithms such as belief propagation (reviewed in [37] and explained in textbook [24]) can compute the maximum *a posteriori* (MAP) for each variable parameter in a factor graph, where the prior probabilities are the initial distributions of the variable nodes, and the constraints on the relationships between variables are encoded as the joint probability distributions at the factors. In our case the initial variable distributions are uniform (i.e., no priors available) and the MAP is the value of the parameters that minimizes the POF objective function. Belief propagation on an acyclic graph is guaranteed to be optimal and efficient, while LBP is a heuristic for cyclic graphs that mimics the message passing of the acyclic belief propagation, except that it provides no guarantee of obtaining the optimal MAP. We use a message passing scheme [36] which has been shown to be empirically successful in computing approximate MAP [38]. In the specification below, $\mu_{X \to f}$ denotes a message from variable node $X$ to factor node $f$ and $\mu_{f \to X}$ denotes a message from variable node $X$ to factor node $f$. Each "message" is a probability distribution, and can be thought of informally as the sender's belief about what the recipient's value should be. During each iteration of message passing, an optimization operation is required. In a joint discrete distribution $g$ with dimension $X_1 X_2,...X_m$, maximization operation over a dimension $X_i$ is defined as

$$\max_{\sim X_i} g(x_1, x_2, ..., x_m) = \sum_{x_1 \in X_1} \sum_{x_2 \in X_2} \cdot\cdot \sum_{x_{i-1} \in X_{i-1}, x_{i+1} \in X_{i+1}} \cdot\cdot \sum_{x_m \in X_m} g(x_1, x_2, ..., x_m)$$

where $\sim X_i$ denotes the set of all dimensions in $g$ except $X_i$. This results in the LBP algorithm [36], as described in Box 1.

---

**A. Initialization:**
 A.1. Compute lookup joint tables for each factor node
 A.2. Set all variable nodes to uniform distribution
**B. Propagation**: repeat until convergence
 B.1. For each factor node $f$
  B.1.1. For each variable node $X \in N(f)$
   B.1.1.1. Collect $\mu_{X_i \to f}$ : the message from variable node $X_i \in N(f) \setminus \{X\}$ to $f$, which is $p(X_i)$, the current probability distribution of $X_i$
   B.1.1.2. Compute $\mu_{f \to X} = \max_{\sim X} \{ p(f) \cdot \prod_i \mu_{X_i \to f} \}$
   B.1.1.3. Send $\mu_{f \to X}$ to the message history of $X$
 B.2. For each variable node $X$
  B.2.1. Update the distribution of $X$ to $p(X) \cdot \prod_i \mu_{f_i \to X}$, where $\mu_{X_i \to f}$ are stored in the message history of $X$
**C. Output**: compute the MAP probability of each variable node
  $MAP(X) = \arg\max\{p(X)\}$

---

Box 1. LBP algorithm for computing the MAP estimates on factor graph

The LBP algorithm starts by initializing the variable nodes to uniform discrete distribution (Box 1 − step A). At the propagation stage (Box 1 − step B), each iteration consists of one round of message passing from each factor node to all associated variable nodes followed by one round of message passing from each variable node to its neighboring factor nodes. Finally, at the output stage (Box 1 − step C), the MAP probability of each variable node is computed based on its final probability distribution.

Numerical errors may emerge when iteratively performing multiplication of joint tables and discrete distribution because the results may contain numbers that are smaller than the floating point precision and hence rounded off as zeros. This means subsequent multiplication involving this zero value will also return zeros. In order to handle this issue, we represent the values in the joint tables and discrete distribution using their logarithms. By doing this, the product operation needs to be replaced by the summation operation (Box 1 − step B.1.1.2 and B.2.1). As it has been pointed out that normalizing the messages or probability distributions does not affect the final MAP results [35], we also perform normalization in the method so that the messages and beliefs are always valid probability distributions at every iteration.

After some iterations of LBP (Box 1 − step B), the variables are expected to converge to a final distribution. There can be multiple criteria for defining convergence (or termination) in the LBP algorithm. The first criterion used in our method is that no normalized message passed in one iteration differs from the previously iteration by more than a tolerance value. Further, as the LBP algorithm is not guaranteed to converge, and in some cases might oscillate [36], the second criterion is a specific bound on the total number of iterations.

### 4.4 Asymptotic analysis

Asymptotic analysis shows the scalability of LBP for large problems. By using discretization on all variables in the factor graph, the algorithm's asymptotic runtime can be represented in terms of the following measures:

- $T$: threshold for the number of iterations for LBP
- $N_s$: number of species in the pathway
- $N_e$: number of different experimental conditions
- $N_t$: number of timepoints used
- $N_r$: number of reaction rates (variable nodes)
- $B$: number of discrete bins used per reaction rate
- $d$: degree of the factor graph

To determine the run time for step A in Box 1, we need to find the size of each lookup table and the number of lookup tables that need to be computed. Each dimension of the joint table corresponds to one variable node, which is one reaction in the corresponding ODE. Hence the number of dimensions of the joint table, which is the degree of the factor graph, is proportional to the number of reactions in the ODE: table size $\propto B^d$. The factor graph defines one factor node for each ODE, for every experiment and at every timepoint excluding the two end timepoints. Therefore, the number of joint tables is proportional to the number of ODEs (or species), experiments and timepoints. Hence, number of tables $\propto N_s N_e N_t$. Therefore, the required time for step A is $\mathcal{O}(N_s N_e N_t B^d)$.

From step B.1 in Box 1, for every iteration, each factor node iterates through each of the neighboring variable nodes and performs multiplication. This is equivalent to the multiplication operation between a joint tables and a one-dimensional discrete distribution, which takes the run time proportional to the size of the joint table, or $B^d$. Since there are $d$ variables connected to the factor node, the computation requires $dB^d$ operations. Thus the time required for step B.1 is $\mathcal{O}(N_s N_e N_t dB^d)$. From step B.2 in Box 1, for every iteration, each variable node needs to compute its new distribution based on the messages from the factor nodes. As there are $N_r$ variable nodes, the maximum the number of computed messages is $N_s N_e N_t$ and each message has size $B$, so the time required for step B.2 is $\mathcal{O}(N_s N_e N_t N_r B)$. To run on all the factor nodes for $T$ iterations, the time required for step B is $\mathcal{O}(TN_s N_e N_t dB^d + TN_s N_e N_t N_r B)$. Since the run time for step B.1 and B.2 dominates that of step A and step C, the total time complexity of the LBP algorithm is $\mathcal{O}(TN_s N_e N_t dB^d + TN_s N_e N_t N_r B)$.

The asymptotic analysis reveals an interesting property of the method, in which the time complexity only scales poorly with the factor graph degree. This means with a factor graph with bounded degree, the method scales well with respect to the number of species, timepoints and discrete bins. Correspondingly, this means the method scales very well on biological pathways with a bounded number of reactions per species.

## 5 RESULTS

As LBP is not guaranteed to converge, empirical performance of the method must be assessed on a spectrum of problems. We implemented BPPE software to execute the method defined in sections 3-4. BPPE is written in C++, the same language as the Copasi parameter estimation tools [39].

A wide variety of parameter estimation algorithms assert claims of supremacy, and algorithms in the "evolutionary strategies" family are particularly well-reviewed [12]. For a fair comparison, we chose a variety of standard methods:

- Local search: Steepest Descent (SD), Levenberg-Marquardt (LM)
- Global search: Genetic Algorithm (GA), Evolution Strategy using Stochastic Ranking (SRES), Particle Swarm Optimization (PSO)

Furthermore, we propose that BPPE should be followed by a local search method such as LM, because a hybrid approach can correct some of the fine-grained inaccuracies created by the discretization and by the numerically approximate derivatives. Four forms of global and two forms of local optimization produce 8 hybrid methods: BPPE_LM, SRES_LM, PSO_LM, GA_LM, BPPE_SD, SRES_SD, PSO_SD and GA_SD. Note that hybrid methods GA_LM and PSO_LM were proposed by Katare et al. [14]. Rodriguez-Fernandez et al. also proposed the hybrid of SRES with a local optimizer [15, 16]. The implementations of the comparison algorithms were used through the Copasi package (version 4.4, build 26). The objective function for these standard methods, using Copasi default settings, is the weighted SSE as described in equation (1). We also assessed the parameter estimation quality using the species maximum relative error (species MRE), a dimensionless metric defined as

$$\text{Species MRE} = \max_{t,j}\left(\frac{N_t}{\sum_m\left(x_i^{data}(t_m)\right)^2}\cdot\left(x_i^{data}(t_j) - x_i^{sim}(t_j,\vec{k})\right)^2\right)\cdot 100\% \quad (3)$$

Species MRE is an improvement over SSE in this case because MRE does not depend on size of the system, and therefore can evaluate of the ability to match data over a range of network sizes. Note that the weight used in (3) is the inverted mean squared of the data for all timepoints, which makes the measure equivalently sensitive to relative errors in any species regardless of whether the species has high or low concentration.

In addition we measured the median and maximum parameter percentage error, or PPE, between the estimated rate constants and the "correct" nominal values. This is a more stringent criterion to assess a parameter estimation method, because if the SSE objective function has multiple minima, different sets of parameters may match the same set of data equally well. The median and maximum PPE are computed as shown in (4) and (5).

$$\text{Median PPE} = \underset{i}{\text{median}} \left( \frac{\left| k_i^{\text{estimated}} - k_i^{\text{nominal}} \right|}{k_i^{\text{nominal}}} \times 100\% \right) \quad (4)$$

$$\text{Maximum PPE} = \underset{i}{\max} \left( \frac{\left| k_i^{\text{estimated}} - k_i^{\text{nominal}} \right|}{k_i^{\text{nominal}}} \times 100\% \right) \quad (5)$$

Data for these experiments were generated using simulation. The simulation method used was LSODA as implemented in Copasi using default settings. We obtained noisy data by introducing Gaussian error into to the exact data. For instance, if the exact data value was 2 and the noise level was 10%, the 10%-noise data would be drawn from a Gaussian distribution with mean 2 and standard deviation 0.2.

### 5.1 Variable timepoints in a ring network

We first explore the simple question of how many measured timepoints are necessary for good performance. Because the spline approximation error increases with $h$, using more timepoints would be expected to improve accuracy. We constructed a series of simple circular networks. Figure 4 illustrates a chemical diagram of a circular network containing species $X_1, X_2, ..., X_n$.

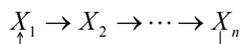

Figure 4. Reaction diagram of a circular network of size $n$

Simulated data are generated with a random set of nominal initial conditions and rate constants. The total time duration was chosen to be 4.0 seconds, and the nominal rate constants were chosen to be the exact bin-midpoints in LBP. Upon performing LBP, we measured the normalized log(POF) value as described below:

$$\text{Normalized log(POF)} = \frac{\log(\text{POF})}{N_t * (N_s - 2)} \quad (6)$$

Log(POF) was normalized with respect to the number of terms in the product formula shown in (2). The term ($N_s$-2) was put in the denominator as there is one factor for each timepoint except two end timepoints. The normalized log(POF) of BPPE with different numbers of timepoints is shown in Figure 5(a). Using 11 timepoints, we also compared the normalized log(POF) value between LBP results and the nominal rate constants, as shown in Figure 5(b).

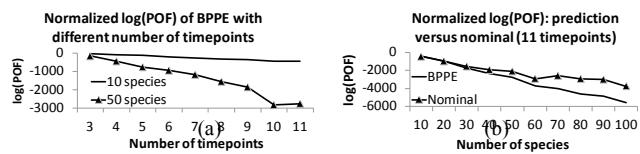

Figure 5. (a) Normalized log(POF) measure on circular networks with respect to the number of timepoints on different network sizes; (b) normalized log(POF) of LBP predicted rate constants versus that of nominal rate constants on circular networks of 10 to 100 species with 11 timepoints

The declining curve in Figure 5(a) indicates that the normalized objective function decreases with more timepoints. This is likely because the approximate derivative is more accurate when the timepoint interval is smaller. Figure 5(b) illustrates that the LBP algorithm minimizes a different objective function, and implies that BPPE chooses the rate constants that have lower POF, even though they may not be the nominal rate constants.

### 5.2 Asymptotic performance on random networks

We first approach the mathematical question of scalability using a series of random low-degree networks with uniformly increasing size from 30 to 140 species. Problems with uniformly increasing size are only possible with an artificial uniform construction. Random binary (A→B) and tertiary (A→B+C or A+B→C) reactions were constructed such that each node has degree at most 3. The number of reactions was chosen to be equal to the number of species, which causes the network to have average degree 2.67.

To evaluate the performance of BPPE_LM, we measured the species MRE, median PPE and maximum PPE of standalone standard methods versus BPPE and BPPE_LM. The tests were performed on noiseless and 20%-noise data. We evaluate the methods on quality of result for a given run time, rather than allowing flexible run time, because the network sizes are large and excessive run time is a practical concern. For the stochastic global search methods (SRES, GA and PSO), we adjusted the run settings such that the run time of all candidates for all standalone methods would be bounded by relatively equal time limits, shown in Figure 7(c).

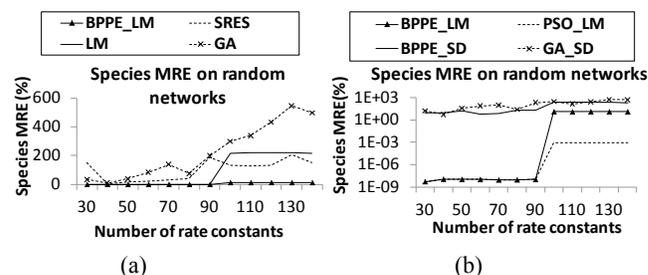

Figure 6. Comparison of species maximum relative error (MRE) on random networks using noiseless data set between: (a) BPPE_LM and standalone methods; and (b) BPPE_LM and hybrid global-local methods

Figure 6(a) compares BPPE_LM with other standalone parameter estimation tools using the noiseless data set. Comparing hybrid against non-hybrid methods is not a fair comparison, and BPPE_LM is predictably superior in this test. Figure 6(b) compares BPPE_LM with other hybrid methods, showing that the hy-brid methods perform equally well in small-sized networks while the PSO_LM method performs best in this test of larger-sized networks. Figure 6(b) also indicates that for these low-degreed networks using noiseless data set, BPPE_LM can match the given data within 10% relative error.

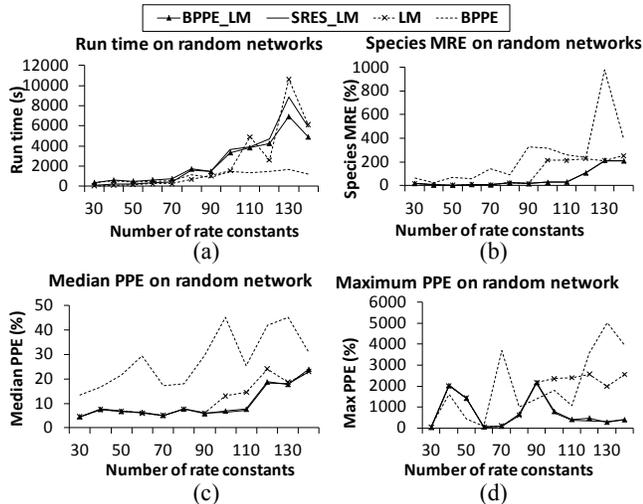

Figure 7. Comparison among parameter estimation approaches on random networks using a 20%-noise data set on: (a) species maximum relative error, (b) run time, (c) median parameter percentage error and (d) maximum parameter percentage error

Four representative methods (BPPE, BPPE_LM, LM and BPPE_LM) were compared on a 20%-noise data set. Figure 7(a) indicates how the run time of each method scales with the number of rate constants. We observe that the BPPE curve has a relatively flat slope, supporting the asymptotic analysis result that BPPE will scale well on networks with bounded degree. In terms of accuracy, (Figure 7(b) shows that BPPE_LM and SRES_LM have almost identical performance where both can match the observed data with relative error of less than 200% (two fold) and Figure 7(c) shows that both can recover the rate constants with a median relative error of less than 25% (Figure 7(c)). However, Figure 7(d) shows that all methods fail to give a reasonable bound on the worst-case percentage error of the estimated rate constants, indicated by the high value of maximum parameter percentage error.

We observe from Figure 6(b) that there is a sharp increase of species MRE between 90 and 100 rate constants in BPPE_LM, PSO_LM and SRES_LM curve. This is caused by the local search nature of LM method, which can get trapped in a local minimum as the network gets more complex with more than 90 rate constants. For smaller networks, the parameter space might be simple enough that LM method found the global minimum and reached near zero matching error and so did all hybrid methods with LM as a post-processor.

The experiments on the random networks indicate that a hybrid approach can perform better than any standalone local or global methods. The assessment of the hybrid methods showed that BPPE_LM SRES_LM, PSO_LM and GA_LM are among the best methods for this type of networks. A complete comparison appears in Supplementary Table S1.

## 5.3 Performance on an Akt activation pathway

In order to test the robustness of our method, we applied our method to modeling the activation dynamics of the survival kinase Akt, in serum-starved mouse embryonic fibroblasts triggered by growth factor. Apart from over-expression of the anti-apoptotic gene bcl-2, the constitutive activation of the pro-survival kinase Akt is one of the most studied survival pathways in tumor cells. In normal cellular functions, Akt phosphorylates and/or interacts with a number of molecules related to cell proliferation, survival, migration and differentiation. Many lines of evidence demonstrate that Akt is a critical player in the development and the progression of tumors. In addition, aberrant hyper-activation of Akt pathway has been detected in up to 50% of all human tumors and is closely associated with chemoresistance (review [40-42]). Therefore, Akt has been an attractive target for anti-cancer drug discovery.

The phosphorylation of Akt is a multi-step process, involving the translocation of Akt from the cytosol to the cell membrane and its phosphorylation by the kinase PDK1 at the Thr308 residue. The translocation of Akt and PDK1 to the membrane has been linked to the amount of phosphatidylinositol 3,4,5-trisphosphate (PIP3) produced at the membrane through the balance between PIP3 production (by the PI3Kinase activated by growth factor) and the degradation of PIP3 by the PTEN phosphatase. Phosphorylation at the $Thr^{308}$ residue is the initial critical step required before a second phosphorylation at $Ser^{473}$ leads to full activation of the kinase before its return into the cytosol. Our newly developed method for parameter estimation was applied to the first step of Akt activation, *e.g.*, the phosphorylation of Akt at $Thr^{308}$. The model was built manually based on our recently published dynamic measurements of Akt phosporylation at $Thr^{308}$, observed in serum-starved mouse embryonic fibroblasts stimulated by the addition of 10% serum to the culture medium. This model includes the production of reactive oxygen species leading to the inactivation of the phosphatase PTEN, in addition to the more conventional pathway of Akt being caused by increased PIP3 production due to induction of the PI3Kinase (PI3K) by serum.

Note that this model includes two simultaneous pathways for serum (growth factors) to influence the PIP2/PIP3 balance: one conventional pathway of PI3K activation, and one under-appreciated pathway in which serum stimulation (growth factors) activates NADPH oxidase complexes (written simply at "NOX" in the diagram) to produce superoxide, which leads to an oxidative inactivation of PTEN (producing PTENox). In the pathway nomenclature, the string "inact" indicates inactive versions of a protein, the suffix "cyto" indicates localization at the cytosol, the suffix "mem" indicates localization at the plasma membrane, and suffix "$p^{308}$" indicates phosphorylation at residue $Thr^{308}$, which is the key event in Akt activation. We applied the BPPE_LM method to a complete dataset generated from the Akt model. The complete reactions are shown in Figure 8, along with the rate parameters of the ODEs as shown in Table 1.

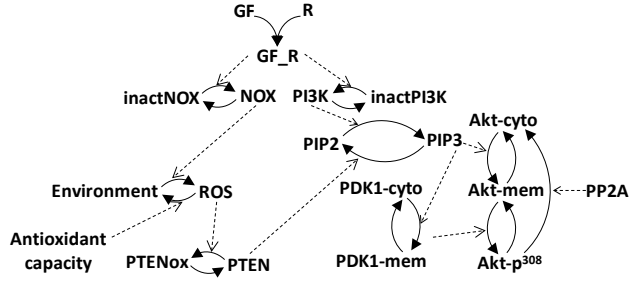

Figure 8. Network diagram for the Akt network, showing the activation of Akt by growth factor (GF) including redox-regulated effects that result in a high level of Akt-p[308].

In Figure 8, dashed arrows represent reactions such as catalysis in which the activating species is not consumed. These reactions are modeled using Michaelis-Menten kinetics or simple enzyme kinetics (catalytic rate × substrate × enzyme).

| Index | Rate constant name | Reaction | Value |
|---|---|---|---|
| 1 | k_uptake | GF + R → GF_R | 14.39 |
| 2 | kM_activNOX | inactNOX → NOX | 0.10 |
| 3 | kcat_activNOX | inactNOX → NOX | 5.19 |
| 4 | kM_activPI3K | inactPI3K → PI3K | 0.09 |
| 5 | kcat_activPI3K | inactPI3K → PI3K | 6.81 |
| 6 | k_deactNOX | NOX → inactNOX | 3.19 |
| 7 | k_deactPI3K | PI3K → inactPI3K | 8.46 |
| 8 | kM_NOX | Environment → ROS | 0.15 |
| 9 | kcat_NOX | Environment → ROS | 13.56 |
| 10 | kM_AntioxidantCapacity | ROS → Environment | 1.00 |
| 11 | kcat_AntioxidantCapacity | ROS → Environment | 50.00 |
| 12 | kM_ROS | PTEN → PTENox | 0.09 |
| 13 | kcat_ROS | PTEN → PTENox | 0.72 |
| 14 | kM_PI3K | PIP2 → PIP3 | 0.30 |
| 15 | kcat_PI3K | PIP2 → PIP3 | 0.40 |
| 16 | kM_PTEN | PIP3 → PIP2 | 0.30 |
| 17 | kcat_PTEN | PIP3 → PIP2 | 0.50 |
| 18 | kcat_PIP3_Akt_cyto | Akt-cyto → Akt-mem | 0.40 |
| 19 | k_Akt_cyto | Akt-mem → Akt-cyto | 0.01 |
| 20 | kcat_PDK1_mem | Akt-mem → Akt-p[308] | 0.60 |
| 21 | kcat_PP2A_Akt_cyto | Akt-p[308] → Akt-cyto | 0.10 |
| 22 | kM_PIP3_PDK1_mem | Akt-p[308] → Akt-mem | 0.50 |
| 23 | kcat_PIP3_PDK1_cyto | PDK1-cyto → PDK1-mem | 0.22 |
| 24 | k_PDK1_mem | PDK1-mem → PDK1-cyto | 0.12 |
| 25 | k_PTEN | PTENox → PTEN | 0.10 |
| 26 | k_Akt_mem | Akt-p[308] → Akt-mem | 0.10 |

Table 1. Rate constant and nominal values in the Akt network. The prefix kM indicates a Michaelis-Mentent constant. The prefix kcat indicates a catalytic rate constant.

Using simulation, we generated data sets with artificial noise at levels of 0%, 1%, and 20%. Using these data sets, we applied 14 parameter estimation methods including 6 standalone methods (BPPE, LM, SD, SRES, PSO and GA) and several promising hybrid methods. The complete comparison of performance results for all methods can be found in Supplementary Table S2.

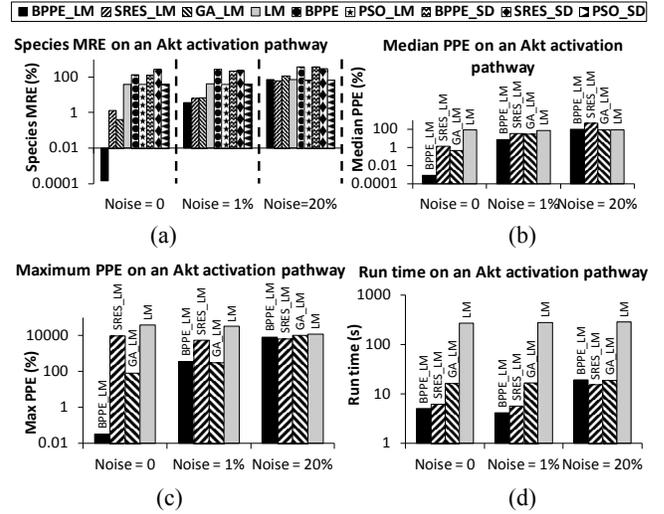

Figure 9. Comparison among different approaches on the Akt network with three noise levels 0, 1% and 20% on: (a) species MRE, (b) median parameter percentage error, (c) maximum parameter percentage error and (d) run time

From Figure 9(a), (b) and (c) it can be observed that the prediction quality of BPPE_LM (BPPE with LM post-processor, see black bar on left) is better than the other methods for noiseless and 1%-noise data set. With noise level of 20%, all methods fail to provide prediction within 100% species maximum relative error, as shown in Figure 9(a). Similarly, on 20%-noise data set, all methods have unacceptably high median and maximum parameter percentage error, as shown in Figure 9(b) and (c).

Figure 9(d) shows an interesting phenomenon, where the LM method takes longer to run than any hybrid method with LM as a post-processor. This indicates that the Akt test case is for some reason difficult for a local search method and when starting with a random initial guess it performs a very large number of iterations before finding a local minimum. However, with a good initial guess provided by a global search method, LM can converge much faster and obtain much better results. In this Akt case the global search result provided by BPPE was a better neighborhood for LM than the other methods, even though BPPE was equivalent on the non-biological tests. A complete comparison among all standalone and hybrid methods and experiment data is shown in Supplementary Table S2.

## 6 CONCLUSION

The chief novelty of the BPPE method is the reformulation of the parameter estimation problem with a factorized objective function, which can be optimized with belief propagation on a factor graph. A direct result of this transformation is that our method searches a tractable number of sparsely connected sub-problems, and it escapes the otherwise universal challenge of generate-and-test search in an exponentially growing space of parameters.

The rate parameters in a biochemical signaling model must be optimized one-at-a-time because changing the value of one parameter can change the behavior of the whole system. Traditional search methods therefore generate a full vector of rate parameters, simulate the model with this full set of parameters, and then accept, reject, or adjust the parameters based on how well the simulation agrees with experimental measurements. This strategy is sampling points on high-dimensional space of parameter vectors (with size exponential in the number of parameters) but for networks with a limited number of unknown parameters, such methods have been empirically successful at finding a basin of convergence with good parameter values. The community has not yet established how large is too large, but the inevitable trend is that networks will eventually have a large enough number of unknown parameters, that a tractable sampling of the parameter space will not be able to explore very many of the basins of convergence. Some form of decomposition [18] will certainly be necessary, and many options are possible.

Our method creates an extreme decomposition, similar to [22], with one differential equation and one timeslice (and one set of experimental conditions) in each sub-problem. BPPE then constructs joint tables of parameter values that satisfy the equation for that timeslice, in a manner similar to a lookup table. Another key point is that the agreement between model and data involves a spline approximation of the species derivative rather than exact comparison with the species concentrations and this can introduce error. The BPPE approach tilts towards asymptotic scalability at the expense of accuracy. We therefore believe that our BPPE method will be less robust to noisy data than simulation-based methods, although the current tests were not able to confirm this belief. In our tests we found that all methods gave unacceptably poor answers with noisy data. Future work must continue to characterize the numerical stability, approximation error, and noise tolerance of this method. The type of modeling problem and other considerations we do not yet understand may eventually prove important for determining what degree of noise, what quality of data, and what degree of discretization the BPPE method can tolerate in realistic biological problems, while still providing meaningful results.

Another distinguishing feature of the BPPE method is that it requires large amounts of concentration measurement data, which would have been prohibitive a decade ago. Recent experimental advances [33, 43] would not only satisfy this constraint, but would be increasingly attractive and cost-effective for studying large networks with many molecules. Previous work in parameter estimation methods assumed that observations would be available only for a sparse subset of proteins in a small signaling network. With mass spectroscopy methods such as SILAC, each timepoint measurement provides simultaneous quantification of the phosphorylation state of all proteins in the system [33, 43]. Under these technologies, there is no additional cost for measuring additional proteins, and model-building strategies can exploit SILAC to scale up the size of the networks under study. This is exactly the future scenario that BPPE_LM can best exploit, with copious data, and where the large number of molecules makes it intractable to sample the high-dimensional parameter space directly. The trade-offs exhibited by our method may be increasingly useful in the future if high-dimensional problems will be estimated in the presence of proteomic measurements.

## ACKNOWLEDGEMENTS

This work was supported by a Lee Kuan Yew Postdoctoral Fellowship and grant R-252-000-342-112 to L.T.-K., by Singapore-MIT Alliance grant C-382-641-004-091 to L.T.-K. and J.K.W, and by grant NMRC/1196/2008 to M.-V.C.